\documentclass[reprint,prl,twocolumn,showpacs,superscriptaddress,amssymb,amsmath,preprintnumbers,floatfix]{revtex4-1}

\usepackage{amsmath}
\usepackage{amssymb}
\usepackage{bm}% bold math
\usepackage{dcolumn}
\usepackage{graphicx}
\usepackage{subfigure}
\usepackage{color}

\begin{document}
%----------------------------------------------------------------
\date{\today}
%----------------------------------------------------------------
\preprint{RIKEN-QHP-26}

%================================================================
\title{%
Complete Tenth-Order QED Contribution to the Muon \boldmath{$g-2$}
}
%----------------------------------------------------------------

\author{Tatsumi Aoyama}
%\email{aoym@riken.jp}
\affiliation{Kobayashi-Maskawa Institute for the Origin
of Particles and the Universe (KMI), Nagoya University,
Nagoya, 464-8602, Japan}
\affiliation{Nishina Center,
RIKEN, Wako, Japan 351-0198 }

\author{Masashi Hayakawa}
%\email{hayakawa@eken.phys.nagoya-u.ac.jp}
%\affiliation{Kobayashi-Maskawa Institute for the Origin of Particles and the Universe (KMI), Nagoya University, Nagoya, 464-8602, Japan}
\affiliation{Department of Physics, Nagoya University, Nagoya, Japan 464-8602 }
\affiliation{Nishina Center,
RIKEN, Wako, Japan 351-0198 }

\author{Toichiro Kinoshita}
%\email{tk@hepth.cornell.edu}
\affiliation{Laboratory for Elementary Particle Physics, Cornell University,
Ithaca, New York, 14853, U.S.A }
\affiliation{Nishina Center,
RIKEN, Wako, Japan 351-0198 }

\author{Makiko Nio}
%\email{nio@riken.jp}
\affiliation{Nishina Center,
RIKEN, Wako, Japan 351-0198 }

%================================================================
\begin{abstract}
 We report the result of our calculation of the complete tenth-order QED
terms of the muon $g-2$.
 Our result is $a_\mu^{(10)} = 753.29~(1.04)$ in units of $(\alpha/\pi)^5$,
which is about 4.5 s.d.~larger than the leading-logarithmic estimate
$663~(20)$.
 We also improved the precision of the eighth-order QED term of $a_\mu$,
obtaining $a_\mu^{(8)} = 130.8794~(63)$ in units of $(\alpha/\pi)^4$.
%Using the best non-QED value of $\alpha$ 
%and the updated fermion mass-ratios,
The new QED contribution is  
$a_\mu(\text{QED}) = 116~584~718~951~(80) \times 10^{-14}$,
which 
%makes the standard-model prediction of $a_\mu$  slightly closer 
%to the measured value, but 
does not resolve the existing discrepancy between
the standard-model prediction and measurement of  $a_\mu$.
%we obtain the standard model prediction 
%$a_\mu(\text{SM}) =
%116~591~840~(59) \times 10^{-11}$,
%to be compared with the measured value
%$a_\mu(\rm{exp}) = 116~592~089~(63) \times 10^{-11}$.
% The difference
%$a_\mu(\text{exp}) - a_\mu(\text{SM}) = 249~(87) \times 10^{-11}$
%is about 2.9 s.d.
\end{abstract}
%================================================================
%


\maketitle
%\tableofcontents

%================================================================

The anomalous magnetic moment $a_\mu$ of the muon 
has been studied extensively both experimentally and theoretically
since it provides one of the promising paths in exploring possible new
physics beyond the standard model.
For this purpose it is crucial to know the prediction of the standard model 
as precisely as possible. 

On the experimental side the current world average of the measured $a_\mu$ is
 \cite{Bennett:2004pv,Roberts:2010cj}:
\begin{equation}
a_\mu(\text{exp}) 
= 116~592~089~(63) \times 10^{-11}~~~[0.5\,\text{ppm}]
\,.
\label{amuexp}
\end{equation}
New experiments designed to improve the precision further
are being prepared at Fermilab  \cite{LeeRoberts:2011zz} and
J-PARC  \cite{Iinuma:2011zz}.

 In the standard model, $a_\mu$ can be divided into
electromagnetic, hadronic, and electroweak contributions
\begin{equation}
a_\mu = a_\mu ({\rm QED}) + a_\mu ({\rm hadronic}) + a_\mu ({\rm electroweak}).
\label{electronanomaly}
\end{equation}
  At present $a_\mu$(hadronic)
is the largest source of theoretical uncertainty.
The uncertainty comes mostly from the 
$O(\alpha^2)$ hadronic vacuum-polarization (v.p.) term,
$\alpha$ being the fine-structure constant.
 The lattice QCD simulations have attempted to evaluate
this contribution
 \cite{Blum:2002ii,Gockeler:2003cw,Aubin:2006xv,
Feng:2011zk,Boyle:2011hu, %DellaMorte:2011ge,
DellaMorte:2011aa}.
 At present, most accurate evaluations must rely on 
the experimental information.
 Three types of measurements are available for this purpose:
(1) $e^{+} e^{-} \rightarrow {\rm hadrons}$,
(2) $\tau^{\pm} \rightarrow \nu + \pi^{\pm} + \pi^0$ ,
(3) $e^{+} e^{-} \rightarrow \gamma + {\rm hadrons}$.
 These processes have been investigated intensely by many groups
 \cite{Davier:2010nc,Jegerlehner:2011ti,Hagiwara:2011af}.%,Jegerlehner:2009ry}.
 We list here one of them \cite{Hagiwara:2011af}:
\begin{equation}
 a_\mu(\rm{had.~v.p.}) = 6949.1~(37.2)_{exp} (21.0)_{rad} \times 10^{-11},
 \label{eq:had_vp_4thOrder}
\end{equation}
which overlaps other values based on  the $e^{+}e^{-}$ data
 \cite{Davier:2010nc,Jegerlehner:2011ti}  
and makes the standard-model  prediction closest 
to the experiment (\ref{amuexp}).
The next-to-leading-order (NLO) hadronic vacuum-polarization contribution
is also known  \cite{Hagiwara:2011af}:
\begin{equation}
 a_\mu(\rm{NLO~had.~v.p.}) = 
 -98.4~(0.6)_{exp} (0.4)_{rad} \times 10^{-11}.
 \label{eq:had_vp_6thOrder}
\end{equation}
The hadronic light-by-light scattering contribution (\textit{l-l}) 
is of similar size
as $a_\mu(\rm{NLO~had.~v.p.})$, but has a much larger theoretical
uncertainty
 \cite{Melnikov:2003xd,Bijnens:2007pz,Prades:2009tw,Nyffeler:2009tw}
\begin{equation}
 a_\mu ({\rm had}.~\text{{\it l-l}}) = 116~(40) \times 10^{-11},
 \label{eq:had_ll}
\end{equation}
where the uncertainty $40\times 10^{-11}$ covers almost all values 
obtained  in different publications.

The electroweak contribution has been calculated up to 2-loop order
 \cite{Fujikawa:1972fe,Czarnecki:1995sz,Knecht:2002hr,Czarnecki:2002nt}:
\begin{equation}
 a_\mu (\rm{weak}) = 154~(2) \times 10^{-11}.
  \label{eq:weak}
\end{equation}
Since this uncertainty is 30 times smaller than the experimental 
precision of (\ref{amuexp}), 
it can be regarded as known precisely.

%The purpose of this paper is to report 
%a new value of the QED contribution  to the standard-model prediction 
%of $a_\mu$.
%The precise value of the QED eighth-order %($\alpha^4$) 
%term 
%is required for a rigarous comparision of the theory and experiment,
%since  its contribution  $ 381 \times 10^{-11}$ 
%is one order of magnitude larger  than 
%the experimental precision in Eq.~(\ref{amuexp}).
%The eighth-order QED term, including the newly evaluated 
%tau-lepton contribution,  is given  with much increased reliability.
%Moreover, the estimate of the QED tenth-order %($\alpha^5$) 
%term  \cite{Kinoshita:2005sm,Kataev:2006yh}
%revealed that its contribution amounts to $4 \times 10^{-11}$,
%which is  the same size of  uncertainties planned in 
%the on-going new experiments  \cite{LeeRoberts:2011zz,Iinuma:2011zz}.
%A complete calculation of the tenth-order QED contribution to $a_\mu$
%is also presented, causing a significant reduction of 
%the theoretical uncertainty of the
%QED part of $a_\mu$. 

%The primary purpose of this letter is to report a sizable reduction of the
%uncertainty of the tenth-order QED term, previously estimated by
%a leading log approximations  \cite{Kinoshita:2005sm,Kataev:2006yh},
%by numerical evaluation of \emph{all} tenth-order terms.
%

The primary purpose of this letter is to report 
the complete numerical evaluation of \emph{all} tenth-order 
QED contribution to $a_\mu$. 
It leads to a sizable reduction of the uncertainty  
of the previous estimate by the leading-log approximations  
 \cite{Kinoshita:2005sm,Kataev:2006yh}.
We have also improved the numerical precision of the eighth-order QED 
contribution including the newly evaluated tau-lepton contribution.
Together they represent a significant reduction in the theoretical
uncertainty of the QED part of $a_\mu$.

% Relevance of individual terms in the perturbative series
%(\ref{amuQED}) may be illustrated by their orders of magnitude;
%%
%\begin{align}
% a_\mu({\rm QED})
% &\simeq 10^{-3} + 4 \times 10^{-5} + 3 \times 10^{-7} \nonumber\\
% &\quad 
% + 4 \times 10^{-9} + 4 \times 10^{-11}
% + O\left(\left(\frac{\alpha}{\pi}\right)^6\right)\,.
%\end{align}
%%
%}
%\textcolor[named]{RoyalBlue}{
%(Or, should more detail values with uncertainties for individual terms 
%be provided after Eq.~(17) ?)
%}
% The purpose of this paper is to report a significant reduction of
%the theoretical uncertainty of $a_\mu({\rm QED})$
%by a complete calculation of the tenth-order QED contribution.
% to $a_\mu$.
%together with the improvement of numerical precision
%of the eighth-order QED contribution.
%}
%(\ref{amuQED})

 The QED contribution to $a_\mu$ can be evaluated
by the perturbative expansion in $\alpha/\pi$:
\begin{equation}
a_\mu ({\rm QED}) = 
\sum_{n=1}^\infty \left( \frac{\alpha}{\pi} \right)^{n} a_\mu^{(2n)},
\label{amuQED}
\end{equation}
where $a_\mu^{(2n)}$ is finite thanks to the renormalizability of QED
and 
can be written as
\begin{align}
 a_\mu^{(2n)} &=
 A_1^{(2n)}+A_2^{(2n)} (m_\mu/m_e)
 + A_2^{(2n)} (m_\mu/m_\tau) \nonumber\\
 &\quad
 + A_3^{(2n)} (m_\mu/m_e ,m_\mu/m_\tau) .
\label{QEDterm}
\end{align}

$A_1^{(2n)}$ is independent of mass 
and universal for all leptons.
 $A_1^{(2)}$, $A_1^{(4)}$ and $A_1^{(6)}$ are known exactly 
 \cite{Schwinger:1948iu,Petermann:1957,Sommerfield:1958,Laporta:1996mq}.
 Mass dependence is known analytically for
$A_2^{(2n)}$ and $A_3^{(2n)}$ for $n=2,3$
 \cite{Samuel:1990qf,Li:1992xf,Laporta:1993ju,Laporta:1992pa,Czarnecki:1998rc}.
We reevaluated them
using the latest values of the muon-electron mass ratio 
$m_\mu/m_e = 206.768~2843~(52)$ 
and/or the muon-tau mass ratio
$m_\mu/m_\tau = 5.946~49~(54) \times 10^{-2}$  \cite{Mohr:2012tt}.
 In the same order of terms as shown on the right-hand-side of (\ref{QEDterm}),
the results are summarized as follows:
\begin{align}
a_\mu^{(2)} &= 0.5,  \nonumber \\
a_\mu^{(4)} &= -0.328~478~965~579~\ldots
               + 1.094~258~312~0~(83) \nonumber \\
            &\quad + 0.780~79~(15) \times 10^{-4}
 \nonumber \\
            &=  0.765~857~425~(17)\,,  \nonumber \\
a_\mu^{(6)} &=  1.181~241~456~\ldots 
                + 22.868~380~04~(23) \nonumber \\
            &\quad + 0.360~70~(13) \times 10^{-3}
                   + 0.527~76~(11) \times 10^{-3}
 \nonumber \\
            &=  24.050~509~96~(32)\,. 
\label{analytic}
\end{align}
%

%-------------------
%
\begin{figure}[t]
\includegraphics[width=7.4cm]{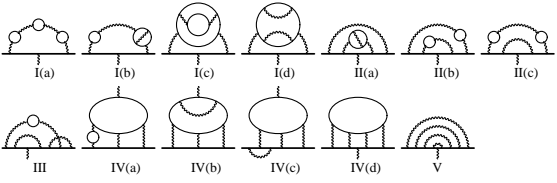}
\caption{
\label{fig:fig8}
 Vertex diagrams representing 13
gauge-invariant subsets contributing to the lepton $g-2$
at the eighth-order.
 Solid and wavy lines represent lepton and photon lines, respectively.
}
\vspace{-0.44cm}
\end{figure}
%
%-------------------

 The value of $a_\mu^{(8)}$ has been obtained mostly
by numerical integration
 \cite{Laporta:1993ds,Kinoshita:2005zr,Aoyama:2007dv,*Aoyama:2007mn}.
 They arise from 13 gauge-invariant sets whose representative diagrams
are shown in Fig.~\ref{fig:fig8}.
 We have reevaluated some of them for further check 
and improvement of numerical precision.
 The results
for the mass-dependent terms
are summarized in Table~\ref{Table:muon8thorder}.

 \begin{table}[b]
 \caption{
 The eighth-order mass-dependent QED contribution 
from 12 gauge-invariant groups to muon $g\!-\!2$,
whose representatives are shown in Fig. \ref{fig:fig8}. 
The mass-dependence of $A_3^{(8)}$ is $A_3^{(8)}(m_\mu/m_e,m_\mu/m_\tau)$.
%%%\text{n.a.}(not available) means that  no contribution to $A_3$ comes 
%%%from a group.
\label{Table:muon8thorder}
 }
 \begin{ruledtabular}
\renewcommand{\baselinestretch}{1.08}
 \begin{tabular}{l@{\hskip-5em}d@{\hskip-5em}d@{\hskip-5em}d}
   group   
&\makebox[-4em]{ $A_2^{(8)}(m_\mu/m_e)$}
&\makebox[-5em]{ $A_2^{(8)}(m_\mu/m_\tau)$}
&\makebox[-2em]{ $A_3^{(8)}$}
\\
\hline
I(a)   &    7.74547~(42)  &    0.000032~(0)   &     0.003209~(0) \\
I(b)   &    7.58201~(71)  &    0.000252~(0)   &     0.002611~(0) \\
I(c)   &    1.624307~(40) &    0.000737~(0)   &     0.001807~(0) \\
I(d)   &   -0.22982~(37)  &    0.000368~(0)   &     0 \\
II(a)  &   -2.77888~(38)  &   -0.007329~(1)   &     0 \\
II(b)  &   -4.55277~(30)  &   -0.002036~(0)   &    -0.009008~(1) \\
II(c)  &   -9.34180~(83)  &   -0.005246~(1)   &    -0.019642~(2) \\
III    &   10.7934~(27)   &    0.04504~(14)   &     0 \\
IV(a)  &  123.78551~(44)  &    0.038513~(11)  &     0.083739~(36) \\
IV(b)  &   -0.4170~(37)   &    0.006106~(31)  &     0 \\
IV(c)  &    2.9072~(44)   &   -0.01823~(11)   &     0 \\
IV(d)  &   -4.43243~(58)  &   -0.015868~(37)  &     0 \\
%\\hline
%\\
%
%sum    &  132.6852~(60)   &    0.04234~(12)   &     0.06272~(4)
 \end{tabular}
\renewcommand{\baselinestretch}{1.00}
 \end{ruledtabular}
 \end{table}

 From the data listed in Table~\ref{Table:muon8thorder} and the value
of $A_1^{(8)}$ from 
Refs.~\cite{Kinoshita:2005zr,Aoyama:2007dv,*Aoyama:2007mn,ae10:PRL},
we obtain the following value for 
the eighth-order QED contribution $a_\mu^{(8)}$:
\begin{align}
a_\mu^{(8)} &= -1.9106~(20) + 132.685~2~(60)   \nonumber \\ 
            &\quad 
               + 0.042~34~(12) + 0.062~72~(4)  \nonumber \\
            &=  130.879~6~(63).  
\label{amu8} 
\end{align}
%

%-------------------
\begin{figure}[t]
\includegraphics[width=7.4cm]{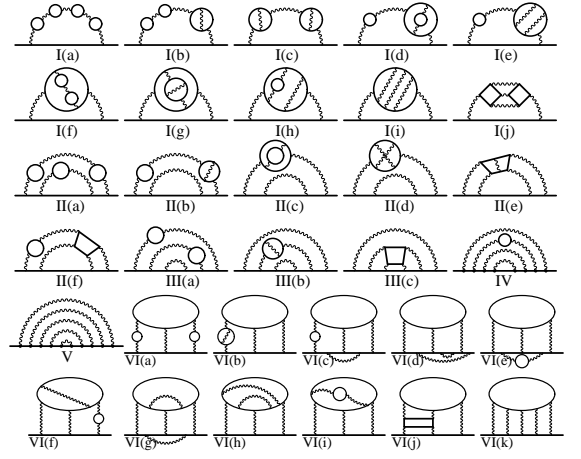}
\caption{
\label{fig:X10}
Self-energy-like diagrams representing 32
gauge-invariant subsets contributing to the lepton $g-2$
at the tenth order.
Solid lines represent lepton lines propagating in a
weak magnetic field.}
\vspace{-0.44cm}
\end{figure}
%
%-------------------

 Over the period of more than nine years we have numerically 
evaluated all 32 gauge-invariant sets of diagrams 
that contribute to $a_\mu^{(10)}$ 
 \cite{Kinoshita:2005sm,Aoyama:2008gy,Aoyama:2010yt,
Aoyama:2008hz,*Aoyama:2010pk,*Aoyama:2011rm,*Aoyama:2010zp,*Aoyama:2011zy,*Aoyama:2011dy,*Aoyama:2012fc,ae10:PRL},
whose representative diagrams are shown in Fig.~\ref{fig:X10}.
The results
for mass-dependent terms
are summarized in Table \ref{Table:10th-order-mass-dependent}.
Some simple diagrams were evaluated analytically or in the asymptotic expansion in $m_\mu/m_e$ \cite{Kataev:1991cp,*Kataev:1991cpErratum,Broadhurst:1992za,Laporta:1994md,Aguilar:2008qj,Baikov:2008si}.
The results are consistent with our numerical ones.

 From the data listed in this Table and the value
of $A_1^{(10)}$ from Ref.~\cite{ae10:PRL}, we obtain
the complete tenth-order result:
\begin{align}
a_\mu^{(10)} &= 9.168~(571) + 742.18~(87)   
%%%%% \nonumber \\ &\quad
               -  0.068~(5) + 2.011~(10)  \nonumber \\
             &= 753.29~(1.04).
\label{amu10}
\end{align}
 The uncertainty $1.04$ is attributed entirely to the statistical 
fluctuation in the Monte-Carlo integration of Feynman amplitudes
by VEGAS  \cite{Lepage:1977sw}. 
This is 20 times more precise than the previous estimate,
$663~(20)$, obtained in the leading-logarithmic 
approximation  \cite{Kinoshita:2005sm}. 
This is mainly because we had underestimated the magnitude of
the contribution of the Set III(a).
 Note also that (\ref{amu10}) 
is about $4.5$ s.d. larger than the leading-log estimate.
 The numerical values of $(\alpha/\pi)^{(n)} a_\mu^{(2n)}$ for
$n=1,2,\cdots,5$  are   summarized in Table~\ref{table:QED}.

\begingroup
%\squeezetable
\begin{table}
\caption{
Tenth-order mass-dependent contribution to the muon $g-2$
from 31 gauge-invariant subsets shown in Fig.~\ref{fig:X10}.
The mass-dependence of $A_3^{(10)}$ is $A_3^{(10)}(m_\mu/m_e,m_\mu/m_\tau)$.
%%%% of Set I, Set II, Set III, Set IV, and Set V.
%%% $n_F$ is the number of vertex 
%%%diagrams contributing to $A_1^{(10)}$.
 \label{Table:10th-order-mass-dependent}
}
\begin{ruledtabular}
%%%%%%\renewcommand{\baselinestretch}{1.08}
%%\begin{tabular}{@{\hskip+2em}l@{\hskip-6em}d@{\hskip-6em}d@{\hskip-6em}d}
\begin{tabular}{l@{\hskip-6em}d@{\hskip-6em}d@{\hskip-6em}d}
   set  &  
\makebox[-5em]{ $A_2^{(10)}(m_\mu/m_e)$ } & 
\makebox[-5em]{ $A_2^{(10)}(m_\mu/m_\tau)$ } & 
%%%\makebox[-5em]{ $A_3^{(10)}$ } & 
\makebox[-5em]{ $A_3^{(10)}$ } 
%%%reference
\\
\hline
I(a)   &   22.566~973~(3)     &    0.000~038~(0)   &     0.017~312~(1) \\
I(b)   &   30.667~091~(3)     &    0.000~269~(0)   &     0.020~179~(1)\\
I(c)   &    5.141~395~(1)     &    0.000~397~(0)   &     0.002~330~(0)\\
I(d)   &    8.8921~(11)       &    0.000~388~(0)   &     0.024~487~(2)\\
I(e)   &   -0.9312~(24)       &    0.000~232~(0)   &     0.002~370~(0)\\
I(f)   &    3.685~049~(90)    &    0.002~162~(0)   &     0.023~390~(2)\\
I(g)   &    2.607~87~(72)     &    0.001~698~(0)   &     0.002~729~(1)\\
I(h)   &   -0.5686~(11)       &    0.000~163~(1)   &     0.001~976~(3)\\
I(i)   &    0.0871~(59)       &    0.000~024~(0)   &     0 \\
I(j)   &   -1.263~72~(14)     &    0.000~168~(1)   &     0.000~110~(5)\\
II(a)  &  -70.4717~(38)       &   -0.018~882~(8)   &    -0.290~853~(85)\\
II(b)  &  -34.7715~(26)       &   -0.035~615~(20)  &    -0.127~369~(60)\\
II(c)  &   -5.385~75~(99)     &   -0.016~348~(14)  &    -0.040~800~(51)\\
II(d)  &    0.4972~(65)       &   -0.007~673~(14)  &     0 \\
II(e)  &    3.265~(12)        &   -0.038~06~(13)   &     0 \\
II(f)  &  -77.465~(12)        &   -0.267~23~(73)   &    -0.502~95~(68)\\
III(a) &  109.116~(33)        &    0.283~000~(32)  &     0.891~40~(44)\\
III(b) &   11.9367~(45)       &    0.143~600~(10)  &     0 \\
III(c) &    7.37~(15)         &    0.1999~(28)     &     0 \\
IV     &  -38.79~(17)         &   -0.4357~(25)     &     0 \\
VI(a)  &  629.141~(12)        &    0.246~10~(18)   &     2.3590~(18)\\
VI(b)  &  181.1285~(51)       &    0.096~522~(93)  &     0.194~76~(26)\\
VI(c)  &  -36.58~(12)         &   -0.2601~(28)     &    -0.5018~(89)\\
VI(d)  &   -7.92~(60)         &    0.0818~(17)     &     0 \\
VI(e)  &   -4.32~(14)         &   -0.035~94~(32)   &    -0.1122~(24)\\
VI(f)  &  -38.16~(15)         &    0.043~47~(85)   &     0.0659~(31)\\
VI(g)  &    6.96~(48)         &   -0.044~51~(96)   &     0 \\
VI(h)  &   -8.55~(23)         &    0.004~85~(46)   &     0\\
VI(i)  &  -27.34~(12)         &   -0.003~45~(33)   &    -0.0027~(11)\\
VI(j)  &  -25.505~(20)        &   -0.011~49~(33)   &    -0.016~03~(58)\\
VI(k)  &   97.123~(62)        &    0.002~17~(16)   &     0 \\

%%%\\hline
%\\
%
%sum    &  742.18~(87)          &   -0.068~(5)     &     2.011~(10) 
\end{tabular}
\renewcommand{\baselinestretch}{1.00}
\end{ruledtabular}
\end{table}
\endgroup

 \begin{table}[b]
 \caption{
Contributions to  muon $g\!-\!2$ from 
%each perturbation terms of QED
QED perturbation term $a_\mu^{(2n)} (\alpha/\pi)^{n} \times 10^{11}$. 
%for $n=1,2,\cdots,5$.
They are evaluated with  two values of the fine-structure constant
determined by the Rb experiment and by the electron $g-2$ ($a_e$).
\label{table:QED}
 }
 \begin{ruledtabular}
\renewcommand{\baselinestretch}{1.08}
 \begin{tabular}{c@{\hskip-7em}d@{\hskip-7em}d}
 \makebox[+5em]{ order  } 
&\makebox[-2em]{ with $\alpha^{-1}(\text{Rb})$ }
&\makebox[-2em]{ with $\alpha^{-1}(a_e)$ }
\\
\hline
%$a_\mu^{(2)} (\alpha/\pi)$   
$2$
             &  116~140~973.318~(77) 
             &  116~140~973.212~(30) 
\\
%$a_\mu^{(4)} (\alpha/\pi)^2  $  
$4$
             &  413~217.6291~(90)  
             &  413~217.6284~(89)  
\\
%$a_\mu^{(6)} (\alpha/\pi)^3  $  
$6$
             &  30~141.902~48~(41)  
             &  30~141.902~39~(40)  
\\
%$a_\mu^{(8)} (\alpha/\pi)^4  $  
$8$
             &  381.008~(19)  
             &  381.008~(19)  
\\
%$a_\mu^{(10)} (\alpha/\pi)^5 $    
$10$
             &  5.0938~(70)  
             &  5.0938~(70)  
\\
$a_\mu(\text{QED}) $          
             &  116~584~718.951~(80)
             &  116~584~718.845~(37)
 
 \end{tabular}
\renewcommand{\baselinestretch}{1.00}
 \end{ruledtabular}
 \end{table}

 In order to evaluate $a_\mu$(QED) using (\ref{amuQED}),
a precise value of $\alpha$ is needed.
 At present,
the best non-QED $\alpha$ is the one obtained from the measurement
of $h/m_\text{Rb}$ \cite{Bouchendira:2010es}, combined with the very precisely known Rydberg constant and
$m_\text{Rb}/m_e$  \cite{Mohr:2012tt}:
\begin{equation}
%\alpha^{-1} (\rm{Rb}) = 137.035~999~037~(91)~~~[0.66~{\rm ppb}].
\alpha^{-1} (\rm{Rb}) =  137.035~999~049~(90)~~~[0.66~{\rm ppb}] .
\label{alphaRb}
\end{equation}
 Actually, we have a more precise value of $\alpha$ which is derived
from the measurement  \cite{Hanneke:2008tm,Hanneke:2010au}
and theory of the electron $g-2$  \cite{ae10:PRL}:
\begin{align}
 \alpha^{-1} (a_e)
 &= 137.035~999~1736~(68)(46)(26)(331)~
  \nonumber\\
 &\quad\  
 [0.25\,{\rm ppb}]\,,
 \label{alphaae}
\end{align}
where the first three uncertainties are 
due to the eighth-order term,
tenth-order term, and the hadronic and electroweak terms,
involved in the evaluation of $a_e$.
 The fourth uncertainty comes from the measurement of $a_e$.
 At present the difference between (\ref{alphaRb}) and
(\ref{alphaae}) is much smaller than the current uncertainty
in the measurement of $a_\mu$ so that 
one may use either one of these two.
%%%\textcolor[named]{RoyalBlue}{
%%% One statement is deleted.
%%% If it is necessary, it will be recovered.
%%%}
%%%
 However, some caution must be exercised
to employ $\alpha^{-1}(a_e)$ to calculate $a_\mu$,
when more accurate experiment of $a_\mu$ becomes available,
because %%%correctness of 
theoretical calculation of $a_e$ 
is strongly correlated with that of $a_\mu$.

  Substituting (\ref{analytic}), (\ref{amu8}), and (\ref{amu10})
in Eq.~(\ref{amuQED}) and using (\ref{alphaRb}), we obtain
\begin{equation}
a_\mu (\rm{QED,Rb}) =
116~584~718~951~(9)(19)(7)(77) \times 10^{-14}\,,
\label{amuQEDRb}
\end{equation}
where  the uncertainties are from 
the lepton mass ratios, the eighth-order term, the tenth-order term,
and the value of $\alpha$ in (\ref{alphaRb}), respectively.
 If we use the value of $\alpha$ in (\ref{alphaae}) instead, we get
\begin{equation}
a_\mu (\text{QED},\,a_e) = 
116~584~718~845~(9)(19)(7)(30) \times 10^{-14}\, .
\label{amuQEDae}
\end{equation}
Note that the uncertainties of the lepton mass ratios, the eighth-order term,
the tenth-order terms, and $\alpha(a_e)$ are  
improved by factors 1.7,~ 1.3, ~20,~ and 1.5, respectively,
compared with  $a_\mu(\text{QED},\,a_e)$ given in Eq.~(99) of Ref.~\cite{Jegerlehner:2009ry}.
%which updates the best value before our work given in %prior to this work
%%
%\begin{equation}
%a_\mu({\rm QED}_{2009},\,a_e) =
%116~584~718~104~(15)(25)(139)(44) \times 10^{-14}\,.
%\label{amuQED:2009}
%\end{equation}
%%
% The %revised 
%large shift of the  value of $a_\mu^{(10)}$ is the dominant source of 
%difference between (\ref{amuQEDae}) and (\ref{amuQED:2009}).

 The difference
between (\ref{amuQEDRb}) and (\ref{amuQEDae})
is less than $1.2 \times 10^{-12}$ so that
we may use either one as far as comparison with the
current experimental data is concerned.

In view of the rather large value of 
$A_2^{(10)} (m_\mu/m_e)$
one might wonder how large $A_2^{(12)}(m_\mu/m_e)$ might be.
As a matter of fact it is not difficult to estimate its size.
For this purpose note that the dominant contribution to
$A_2^{(8)} (m_\mu/m_e)$
comes from the Group IV(a) and the dominant contribution to
$A_2^{(10)}(m_\mu/m_e)$ 
comes from the Set VI(a).
Both are integrals obtained by inserting several second-order
vacuum-polarization loops $\Pi_2$ into the virtual photon lines
of the sixth-order diagram $A_2^{(6)} (m_\mu/m_e; \text{{\it l-l}}\,)$
which contains a light-by-light scattering electron loop.
Analogously the leading contribution to the twelfth-order term
will come from insertion of three $\Pi_2$'s in 
$A_2^{(6)} (m_\mu/m_e; \text{{\it l-l}}\,)$,
namely,
\begin{align}
&A_2^{(12)}(m_\mu/m_e) 
 \sim A_2^{(6)} (m_\mu/m_e; \text{{\it l-l}}\,) 
\nonumber \\
& 
~~~~~~~~~~~~~~~~~~~~~~~
\times \left \{ \frac{2}{3} \ln \left (\frac{m_\mu}{m_e} \right ) 
    - \frac{5}{9} \right\}^3 \times 10 
%%%%\times \left(\frac{\alpha}{\pi}\right)^6
%%%\nonumber \\
\\
\text{and}&  \nonumber \\
&A_2^{(12)}(m_\mu/m_e) \times \left(\frac{\alpha}{\pi}\right)^6
                      \sim 0.8 \times 10^{-12},
\end{align}
noting that $A_2^{(6)} (m_\mu/m_e; \text{{\it l-l}}\,) \sim 20$
and the factor 10 accounts for the possible ways of insertion of $\Pi_2$.
Including the contribution of other diagrams, the size of the 12th-order
term might be as large as $10^{-12}$. 
This is larger than the uncertainty
of the 10th-order term 
in (\ref{amuQEDRb})
so that it would be desirable to obtain at least a crude evaluation
of this term.
%\textcolor[named]{Red}{
% To know its signficance more reliably,
%it is thus desirable to evaluate this term
%in the leading-logarithmic approximation.
%}

Adding 
(\ref{eq:had_vp_4thOrder}),
(\ref{eq:had_vp_6thOrder}),
(\ref{eq:had_ll}), (\ref{eq:weak}),
and (\ref{amuQEDRb}),
and using $\alpha$ from (\ref{alphaRb}),
the theoretical value of $a_\mu$ in the standard model is given by
\begin{equation}
a_\mu (\text{SM}) = 116~591~840~(59) \times 10^{-11}.
\end{equation}
 We have therefore
\begin{equation}
 a_\mu (\text{exp}) - a_\mu (\text{SM}) = 249~(87) \times 10^{-11}.
\end{equation}

The size of discrepancy between theory and experiment has not changed much,
since the tenth-order QED contribution
is not a significant source of theoretical uncertainties.
 Let us emphasize, however, that the 
complete calculation of $a_\mu^{(10)}$ enables us to
concentrate on improving the precision of the hadronic contributions.

\begin{acknowledgments}
We thank J.~Rosner for a helpful comment.
This work is supported in part by the JSPS Grant-in-Aid for
Scientific Research (C)20540261 and (C)23540331.
T.~K.'s work is supported in part by the U. S. National Science Foundation
under Grant No. NSF-PHY-0757868.
T.~K. thanks RIKEN for the hospitality extended to him
while a part of this work was carried out.
Numerical calculations were conducted on 
RSCC and RICC supercomputer systems at RIKEN.
%the
%RIKEN Super Combined Cluster System (RSCC) and the
%RIKEN Integrated Cluster of Clusters (RICC) supercomputing systems.
\end{acknowledgments}
%================================================================

%\bibliographystyle{apsrev}
%\bibliographystyle{plain}
\bibliography{bwoarX}

\end{document}